\documentclass[fleqn,usenatbib]{mnras}
\usepackage{newtxtext,newtxmath}
\usepackage[T1]{fontenc}
\DeclareRobustCommand{\VAN}[3]{#2}
\let\VANthebibliography\thebibliography
\def\thebibliography{\DeclareRobustCommand{\VAN}[3]{##3}\VANthebibliography}
\usepackage{graphicx}	% Including figure files
\usepackage{amsmath}	% Advanced maths commands

\title[Dark particle mass effects on neutron stars]{Dark particle mass effects on neutron star properties from a short-range correlated hadronic model} 

\author[M. Dutra, C. H. Lenzi, and O. Louren\c{c}o]{M. Dutra,\thanks{E-mail: marianad@ita.br}
C. H. Lenzi, and O. Louren\c{c}o
\\
Departamento de F\'isica, Instituto Tecnol\'ogico de Aeron\'autica, DCTA, 12228-900, S\~ao Jos\'e dos Campos, SP, Brazil\\
}

\pubyear{2022}

\begin{document}
\label{firstpage}
% \pagerange{\pageref{firstpage}--\pageref{lastpage}}
\maketitle

\begin{abstract}
In this work we study a relativistic mean-field (RMF) hadronic model, with {nucleonic} short-range correlations (SRC) included, coupled to dark matter (DM) through the Higgs boson. We study different parametrizations of this model by running the dark particle Fermi momentum, and its mass in the range of $50\mbox{ GeV}\leqslant M_\chi\leqslant 500\mbox{ GeV}$, compatible with experimental spin-independent scattering cross-sections. By using this \mbox{RMF-SRC-DM} model, we calculate some neutron star quantities, namely, mass-radius profiles, dimensionless tidal deformabilities, and crustal properties. Our findings show that is possible to construct \mbox{RMF-SRC-DM} parametrizations in agreement with constraints provided by LIGO and Virgo collaboration (LVC) on the GW170817 event, and recent observational data from the NICER mission. Furthermore, we show that the increase of $M_\chi$ favors the model to attain data from LVC regarding the tidal deformabilities. Higher values of $M_\chi$ also induce a reduction of the neutron star crust (mass and thickness), and cause a decrease of the crustal fraction of the moment of inertia ($I_{\mbox{\tiny crust}}/I$). Nevertheless, we show that some \mbox{RMF-SRC-DM} parametrizations still exhibit $I_{\mbox{\tiny crust}}/I>7\%$, a condition that explains the glitch activity in rotation-powered pulsars such as the Vela one. Therefore, dark matter content can also be used for describing such a phenomenon. 
\end{abstract}

\begin{keywords}
Neutron Stars -- Equation of States -- Dark Matter
\end{keywords}

\section{Introduction} 
{A compact star (CS) is a highly dense astrophysical object} composed by hadrons, leptons, and/or quark matter~\citep{lattimer-review1,lattimer-review2,debora-universe,stone-universe,nature-stars}. Therefore, a deep understanding of strong and weak interactions is extremely important in order to correctly describe this system. For the theoretical formulation of the former at the hadronic level, many phenomenological models with a huge amount of parametrizations have been used over the years, such as the Skyrme~\citep{dutra12,skyrmeligo} and relativistic mean-field (RMF) models~\citep{dutra14,rmf-stellar,rmf-stellar-errata,rmfligo}. With the same aim, quark-meson coupling models, in which nucleons are explicitly formed by interacting quarks, are also frequently adopted for nuclear and stellar matter description~\citep{stone-frontiers,stone-review,guichon,mqmc-tobias,cpcnosso}. The {CS} environment can be used to test the interactions allowed by these different models. In that sense, the stellar matter is seen as a natural laboratory, not possible to be reproduced in terrestrial experiments (at least so far), in which the extreme condition of high density is possible to be attained. 

Recently, many observational data were obtained and used as constraints to select the more suitable models and parametrizations, and consequently the related physics, capable of better explaining different {CSs} global properties. As an example of these updated data, we address the reader to the multimessenger astronomy era, started from the detection of gravitational waves originated by the collision of black holes~\citep{bholes1,bholes2,bholes3} and by the {CSs} merging~\citep{ligo17,era,GW190425}. From the latter (GW170817 event), limits on the tidal deformabilities were also inferred~\citep{ligo18}. The x-ray telescope based at the International Space Station, and responsible for NASA's Neutron star Interior Composition Explorer (NICER) mission~\citep{nicer}, also provided valuable information on the mass-radius {relation} of {CSs}, determined from measurements of the pulsars PSR J0030+0451~\citep{0451a,0451b} and PSR J0740+6620~\citep{6620a,6620b}. Furthermore, accurate mass measurements recently led to $M=2.08^{+0.07}_{-0.07}M_{\odot}$ for massive {CSs}, according to~\cite{fonseca}. In summary, there is a lot of recent astrophysical information that can be used to test the limits of hadronic phenomenological models and their different versions. 

The attempt of satisfying all aforementioned observational constraints, and other ones, led to the inclusion of other degrees of freedom beyond baryons, mesons, quarks, and gluons in the formulation of refined hadronic and quark models for the dense matter. An example is the inclusion of dark matter~(DM) in the equations of state furnished as input for the construction of stellar matter. Indications of the existence of this kind of matter were given in the 70's by Vera Cooper Rubin, who verified that the rotational velocity of stars around the center of spiral galaxies ($v_r$) does not obey gravitation theory, which predicts $v(r)\approx \sqrt{M_g/r}$ ($r$ is the distance of the star from the galaxy center, and $M_g$ is the galaxy mass). In~\citep{vera}, rotation curves of a sample of 10 high-luminosity spiral galaxies were analyzed, with the approximate flat format verified for all of them, indicating that $v(r)$ is roughly constant. From the previous expression for $v(r)$, one observes that such a result is only possible if $M_g$ increases proportionally to~{$r$}, i.e, the mass of the galaxies should be higher than the observed visible matter (they would contain massive halos). This non detected mass, named dark matter, was earlier suggested by astronomer Fritz Zwicky~\citep{zwicky-original,zwicky}. His measurements and calculations for the Coma cluster led him to increase the mass of each galaxy of the cluster by two orders of magnitude, indicating that it is not composed only of luminous matter. Similar results were also found by Jan Hendrik Oort~\citep{oort}, who studied the gravitational force perpendicular to the galactic plane. Despite the research on this subject presents many evidences of the existence of DM, its exact nature is not yet completely understood. There are different types of possible DM candidates, such as weakly interacting massive particles (WIMP), sterile neutrinos, axinos, gravitinos, axions, Q-balls, WIMPzillas, supersymmetric, and mirror matter~\citep{cand1,cand2}. Recent studies have been developed in which dark particle candidates are coupled to hadrons in order to describe {CSs} properties, see for instance~\cite{rmfdm13,rmfdm1,rmfdm2,rmfdm3,abdul,rmfdm6,rmfdm11,rmfdm10,rmfdm8,rmfdm7,rmfdm12}. 

{CSs are suitable environments for DM binding because they are extremely massive systems (around $2M_\odot\approx 10^{30}$~kg) and, therefore, gravitational interaction is very intense. Other type of interactions between DM and observable matter are also possible, since some upper limits are verified.} In particular, in~\cite{dmnosso1,dmnosso2} the lightest neutralino with mass of $M_\chi=200$~GeV was used as dark particle coupled through the Higgs boson to nucleons presenting short-range correlations (SRC)~\citep{sciencesrc1,naturesrc2,naturescr3,naturesrc4,hen2017,duer2019,rev3,cai,baoanli21,baoanli22,lucas}. Such correlations are experimentally verified in electron-induced quasielastic proton knockout reactions, in which nucleons correlate in pairs with high relative momentum. The effect of this phenomenology is observed in the single-nucleon momentum distribution, $n(k)$, in finite nuclei and nuclear matter, where a high momentum tail of the form $k^{-4}$ takes place. This new structure for $n(k)$ changes all kinetic terms of the hadronic model and, consequently, all related thermodynamics even at the zero temperature regime, used to construct the equations of state for the stellar matter description. In this work, we use this particular model coupled to dark matter (\mbox{RMF-SRC-DM} model) to calculate {CSs} global properties, namely, mass-radius diagrams, dimensionless tidal deformabilities related to the GW170817 event, and also some specific properties of the crust such as mass, thickness, and the fraction of the moment of inertia. Unlike in~\cite{dmnosso1,dmnosso2} in which the dark particle mass was kept fixed in $M_\chi=200$~GeV, we now let this quantity to run inside the range of $50\mbox{ GeV}\leqslant M_\chi\leqslant 500\mbox{ GeV}$. Our results point out the possibility of generating \mbox{RMF-SRC-DM} parametrizations, with different values of $M_\chi$, capable of reproducing observational data from the GW170817 event, and the recent findings of the NICER mission. Concerning the crustal properties, we also show that the increase of $M_\chi$ reduces both, the {CS} crust region and also a crustal fraction of the moment of inertia. Even with the latter effect, we show that this quantity can attain a value greater than $7\%$. As this particular constraint is used to explain the glitches observed in pulsars such as the Vela one, it can not be disregarded the possibility of this phenomenon being induced by dark matter. 

This paper is divided as follows: In Sec.~\ref{form} we present the formalism of coupling between DM and the hadronic model with short-range correlations included. The Lagrangian density and the main thermodynamical equations of state are also shown in this section. Sec.~\ref{results} contains our results. More specifically, in Sec.~\ref{resultsmr} mass-radius profiles and dimensionless tidal deformabilities are tested against the variation in both, $M_\chi$ and the dark particle Fermi momentum. Crustal properties are calculated in Sec.~\ref{resultscrust}. Finally, we close the manuscript with a summary and concluding remarks in Sec.~\ref{summary}.

\section{Dark Matter-Hadronic model with short-range correlations} 
\label{form}

The model used in this work considers nucleons (protons/neutrons), with short-range correlations included,  interacting with each other through mesons exchange and with a certain dark matter content. The hadronic part of the system is described by the following Lagrangian density
\begin{align}
&\mathcal{L}_{\mbox{\tiny HAD}} = \overline{\psi}(i\gamma^\mu\partial_\mu - M_{\mbox{\tiny nuc}})\psi 
+ g_\sigma\sigma\overline{\psi}\psi 
- g_\omega\overline{\psi}\gamma^\mu\omega_\mu\psi
\nonumber \\ 
&- \frac{g_\rho}{2}\overline{\psi}\gamma^\mu\vec{\rho}_\mu\vec{\tau}\psi
+\frac{1}{2}(\partial^\mu \sigma \partial_\mu \sigma - m^2_\sigma\sigma^2)
- \frac{A}{3}\sigma^3 - \frac{B}{4}\sigma^4 
\nonumber\\
&-\frac{1}{4}F^{\mu\nu}F_{\mu\nu} 
+ \frac{1}{2}m^2_\omega\omega_\mu\omega^\mu 
+ \frac{C}{4}(g_\omega^2\omega_\mu\omega^\mu)^2 -\frac{1}{4}\vec{B}^{\mu\nu}\vec{B}_{\mu\nu} 
\nonumber \\
&+ \frac{1}{2}\alpha'_3g_\omega^2 g_\rho^2\omega_\mu\omega^\mu
\vec{\rho}_\mu\vec{\rho}^\mu + \frac{1}{2}m^2_\rho\vec{\rho}_\mu\vec{\rho}^\mu.
\label{dlag}
\end{align}
The fields of this theory, namely, $\psi$, $\sigma$, $\omega^\mu$, and $\vec{\rho}_\mu$, represent the nucleon and the mediators (mesons) $\sigma$, $\omega$, and $\rho$, with the respective masses given by $M_{\mbox{\tiny nuc}}$, $m_\sigma$, $m_\omega$, and $m_\rho$. The coupling constants $g_\sigma$, $g_\omega$, $g_\rho$, $A$, $B$, {$C$} and $\alpha'_3$ control the strength of each interaction, and the tensors read $F_{\mu\nu}=\partial_\nu\omega_\mu-\partial_\mu\omega_\nu$ and $\vec{B}_{\mu\nu}=\partial_\nu\vec{\rho}_\mu-\partial_\mu\vec{\rho}_\nu$. 

In order to describe the entire system with dark matter interacting with hadrons, we adopt the same Lagrangian density used in~\cite{dmnosso1,dmnosso2}, namely,
\begin{align}
\mathcal{L} &= \overline{\chi}(i\gamma^\mu\partial_\mu - M_\chi)\chi
+ \xi h\overline{\chi}\chi +\frac{1}{2}(\partial^\mu h \partial_\mu h - m^2_h h^2)
\nonumber\\
&+ f\frac{M_{\mbox{\tiny nuc}}}{v}h\overline{\psi}\psi + \mathcal{L}_{\mbox{\tiny HAD}},
\label{dlagtotal}
\end{align}
in which the dark fermion is represented by the Dirac field $\chi$ with mass $M_\chi$. The interaction between $\chi$ and $\psi$ is due to the Higgs boson (scalar field $h$) with mass $m_h=125$~GeV, and its strength is regulated by the constant $fM_{\mbox{\tiny nuc}}/v$, where $v=246$~GeV is the Higgs vacuum expectation value. The constant $\xi$ is related to the Higgs-dark particle coupling. Here we use $\xi=0.01$~\citep{dmnosso1,dmnosso2}, and the central value $f=0.3$ obtained in~\cite{cline,clineerrata}. According to~\cite{coolingdm}, these values ensure a spin-independent scattering cross-sections in agreement with experimental data from PandaX-II~\citep{panda}, LUX~\citep{lux}, and DarkSide~\citep{darkside} collaborations for dark fermion mass in the range of $50\mbox{ GeV}\leqslant M_\chi\leqslant 500\mbox{ GeV}$.

The use of the mean-field approximation in this theory leads to $\sigma\rightarrow \left<\sigma\right>\equiv\sigma$, $\omega_\mu\rightarrow \left<\omega_\mu\right>\equiv\omega_0$, $
\vec{\rho}_\mu\rightarrow \left<\vec{\rho}_\mu\right>\equiv \bar{\rho}_{0(3)}$, and $h\rightarrow \left<h\right>\equiv h$. This procedure is useful to determine the field equations, namely,
\begin{align}
m^2_\sigma\,\sigma &= g_\sigma\rho_s - A\sigma^2 - B\sigma^3 
\\
m_\omega^2\,\omega_0 &= g_\omega\rho - Cg_\omega(g_\omega \omega_0)^3 
- \alpha_3'g_\omega^2 g_\rho^2\bar{\rho}_{0(3)}^2\omega_0, 
\\
m_\rho^2\,\bar{\rho}_{0(3)} &= \frac{g_\rho}{2}\rho_3 
-\alpha_3'g_\omega^2 g_\rho^2\bar{\rho}_{0(3)}\omega_0^2, 
\\
[\gamma^\mu (&i\partial_\mu - g_\omega\omega_0 - g_\rho\bar{\rho}_{0(3)}\tau_3/2) - M^*]\psi = 0,
\\
m^2_h\,h &= \xi\rho_s^{\mbox{\tiny DM}} + f\frac{M_{\mbox{\tiny nuc}}}{v}\rho_s
\\
(\gamma^\mu &i\partial_\mu - M_\chi^*)\chi = 0,
\end{align}
with $\tau_3=1$ ($-1$) for protons (neutrons). The effective nucleon and dark fermion masses are 
\begin{eqnarray}
M^* = M_{\mbox{\tiny nuc}} - g_\sigma\sigma - f\frac{M_{\mbox{\tiny nuc}}}{v}h
\end{eqnarray}
and 
\begin{eqnarray}
M^*_\chi = M_\chi - \xi h,
\end{eqnarray}
respectively, and the densities are $\rho_s=\left<\overline{\psi}\psi\right>={\rho_s}_p+{\rho_s}_n$, $\rho=\left<\overline{\psi}\gamma^0\psi\right> = \rho_p + \rho_n$, $\rho_3=\left<\overline{\psi}\gamma^0{\tau}_3\psi\right> = \rho_p - \rho_n=(2y_p-1)\rho$, and
$\rho_s^{\mbox{\tiny DM}} = \left<\overline{\chi}\chi\right>$, with
\begin{eqnarray}
\rho_s^{\mbox{\tiny DM}} &=& 
\frac{\gamma M^*_\chi}{2\pi^2}\int_0^{k_F^{\mbox{\tiny DM}}} \hspace{-0.5cm}\frac{k^2dk}{(k^2+M^{*2}_\chi)^{1/2}}.
\end{eqnarray}
$\gamma=2$ is the degeneracy factor, and the proton fraction is denoted by $y_p=\rho_p/\rho$, with nucleon densities given by $\rho_{p,n}=\gamma{k_F^3}_{p,n}/(6\pi^2)$. The Fermi momenta related to protons/neutrons, and the dark particle, are ${k_F}_{p,n}$ and $k_F^{\mbox{\tiny DM}}$, respectively. 

From the Lagrangian density of the system, it is possible to determine the energy-momentum tensor ($T^{\mu\nu}$), and from this quantity we obtain energy density and pressure. In our case, Eq.~\ref{dlagtotal} is used to calculate $\mathcal{E}=\left<T_{00}\right>$ and $P=\left<T_{ii}\right>/3$. This procedure gives rise to the following expressions,
\begin{align} 
&\mathcal{E} = \frac{m_{\sigma}^{2} \sigma^{2}}{2} +\frac{A\sigma^{3}}{3} +\frac{B\sigma^{4}}{4} 
-\frac{m_{\omega}^{2} \omega_{0}^{2}}{2} - \frac{Cg_{\omega}^4\omega_{0}^4}{4}
- \frac{m_{\rho}^{2} \bar{\rho}_{0(3)}^{2}}{2} 
\nonumber\\
&+ g_{\omega} \omega_{0} \rho + \frac{g_{\rho}}{2} 
\bar{\rho}_{0(3)} \rho_{3}   -\frac{1}{2} \alpha'_3 g_{\omega}^{2} g_{\rho}^{2} \omega_{0}^{2} 
\bar{\rho}_{0(3)}^{2} + \mathcal{E}_{\mathrm{kin}}^{p} + \mathcal{E}_{\mathrm{kin}}^{n}
\nonumber\\
&+ \frac{m_h^2h^2}{2} + \mathcal{E}_{\mathrm{kin}}^{\mbox{\tiny DM}},
\label{eden}
\end{align}
and
\begin{align}
&P = -\frac{m_{\sigma}^{2} \sigma^{2}}{2} - \frac{A\sigma^{3}}{3} - \frac{B\sigma^{4}}{4} 
+ \frac{m_{\omega}^{2} \omega_{0}^{2}}{2} + \frac{Cg_{\omega}^4\omega_0^4}{4}
\nonumber\\
&+ \frac{m_{\rho}^{2} \bar{\rho}_{0(3)}^{2}}{2} + \frac{1}{2} \alpha'_3 g_{\omega}^{2} 
g_{\rho}^{2} \omega_{0}^{2} \bar{\rho}_{0(3)}^{2} + P_{\mathrm{kin}}^{p} + P_{\mathrm{kin}}^{n}
- \frac{m_h^2h^2}{2} 
\nonumber\\
&+ P_{\mathrm{kin}}^{\mbox{\tiny DM}}.
\label{press}
\end{align}
The dark particle kinetic terms are
\begin{eqnarray}
\mathcal{E}_{\mbox{\tiny kin}}^{\mbox{\tiny DM}} &=& \frac{\gamma}{2\pi^2}\int_0^{k_F^{\mbox{\tiny DM}}}\hspace{-0.3cm}k^2(k^2+M^{*2}_\chi)^{1/2}dk,\qquad\mbox{and}
\label{ekindm}
\\
P_{\mbox{\tiny kin}}^{\mbox{\tiny DM}} &=& 
\frac{\gamma}{6\pi^2}\int_0^{{k_F^{\mbox{\tiny DM}}}}\hspace{-0.5cm}\frac{k^4dk}{(k^2+M^{*2}_\chi)^{1/2}}.
\label{pkindm}
\end{eqnarray}

Concerning the hadronic part of the RMF-DM model, the inclusion of the SRC is performed by replacing the usual step functions present in kinetic terms by the one including the high momentum tail~\citep{cai,lucas} given by $n_{n,p}(k) = \Delta_{n,p}$ for $0<k<k_{F\,{n,p}}$, and $n_{n,p}(k) = C_{n,p}\,(k_{F\,{n,p}}/k)^4$ for $k_{F\,{n,p}}<k<\phi_{n,p} \,k_{F\,{n,p}}$ in which $\Delta_{n,p}=1 - 3C_{n,p}(1-1/\phi_{n,p})$, $C_p=C_0[1 - C_1(1-2y_p)]$, $C_n=C_0[1 + C_1(1-2y_p)]$, $\phi_p=\phi_0[1 - \phi_1(1-2y_p)]$ and $\phi_n=\phi_0[1 + \phi_1(1-2y_p)]$. Here we use $C_0=0.161$, $C_1=-0.25$, $\phi_0 = 2.38$ and $\phi_1=-0.56$~\citep{cai,lucas}. It leads to the following modified kinetic terms,
\begin{eqnarray} 
\mathcal{E}_{\text {kin }}^{n,p} &=& \frac{\gamma \Delta_{n,p}}{2\pi^2} \int_0^{{k_{F\,{n,p}}}} 
k^2dk({k^{2}+M^{* 2}})^{1/2}
\nonumber\\
&+& \frac{\gamma C_{n,p}}{2\pi^2} \int_{k_{F\,{n,p}}}^{\phi_{n,p}\, {k_{F\,{n,p}}}} 
\frac{{k_F}_{n,p}^4}{k^2}\, dk({k^{2}+M^{* 2}})^{1/2},
\nonumber \\
P_{\text {kin }}^{n,p} &=&  
\frac{\gamma \Delta_{n,p}}{6\pi^2} \int_0^{k_{F\,{n,p}}}  
\frac{k^4dk}{\left({k^{2}+M^{*2}}\right)^{1/2}} 
\nonumber\\
&+& \frac{\gamma C_{n,p}}{6\pi^2} \int_{k_{F\,{n,p}}}^{\phi_{n,p}\, {k_{F\,{n,p}}}} 
 \frac{{k_F}_{n,p}^4dk}{\left({k^{2}+M^{*2}}\right)^{1/2}},
\end{eqnarray}
and modified scalar densities 
\begin{align}
&{\rho_s}_{n,p} = 
\frac{\gamma M^*\Delta_{n,p}}{2\pi^2} \int_0^{k_{F\,{n,p}}}  
\frac{k^2dk}{\left({k^{2}+M^{*2}}\right)^{1/2}} 
\nonumber\\
&+ \frac{\gamma M^*C_{n,p}}{2\pi^2} \int_{k_{F\,{n,p}}}^{\phi_{n,p}\, {k_{F\,{n,p}}}} 
\frac{{k_F}_{n,p}^4}{k^2}  \frac{dk}{\left({k^{2}+M^{*2}}\right)^{1/2}}
\end{align}
for protons and neutrons.

\section{Results} 
\label{results}

In this section, we present the predictions of the \mbox{RMF-SRC-DM} model regarding neutron stars properties. Specifically for the hadronic side of the system, we focus on the parametrization discussed in~\cite{piekarewicz2014}, namely, TFcmax. The authors constructed RMF parametrizations with the aim of maximizing the transition pressure ($p_t$) at the core-curst interface, the quantity shown to be important, for instance, for the description of the pulsar Vela glitches, since the crustal moment of inertia fraction is directly affected by $p_t$. Furthermore, all these studied parametrizations also provide a good description of various finite nuclei properties, according to~\cite{piekarewicz2014}. Concerning the bulk properties at the saturation density ($\rho_0$) given by the TFcmax parametrization, we list the following ones: $\rho_0=0.148$~fm$^{-3}$, $B_0=-16.47$~MeV (binding energy), $M^*_0/M_{\mbox{\tiny nuc}}=0.57$ (effective nucleon mass ratio), $K_0=260$~MeV (incompressibility), $\tilde{J}\equiv\mathcal{S}(2\rho_0/3)=29.4$~MeV (symmetry energy at $2\rho_0/3$) and $L_0=74$~MeV (symmetry energy slope). It is worth mentioning that this last quantity is inside of two particular ranges, namely, $L_0=(106\pm37)$~MeV, pointed out in~\cite{piekaprex2} to be consistent with results of neutron skin thickness measurements of $^{208}\rm Pb$~\citep{prex2}, and $42\leqslant L_0\leqslant 117$~MeV, obtained from the analysis of charged pions spectra~\citep{pions}.

We emphasize that these bulk parameters were imposed in the hadronic model used here. Since its structure contains the modifications provided by the inclusion of the SRC phenomenology, the coupling constants of the \mbox{RMF-SRC} model must be different from those shown in~\cite{piekarewicz2014} to keep the same bulk quantities. This is the case in our study, i.e., we use here a SRC version of the TFcmax model with the same bulk parameters of the original parametrization, namely, the aforementioned ones. It was also shown in~\citep{dmnosso1} that the presence of SRC favors the inclusion of DM in the system since the increase of the maximum neutron star mass caused by this phenomenology (inclusion of SRC) balances the decrease induced by DM.

With regard to the dark matter side of the \mbox{RMF-SRC-DM} model, we explore the variation of the dark particle mass and its effects on the neutron star properties. In~\cite{dmnosso1} and~\cite{dmnosso2}, the lightest neutralino was used as the dark matter candidate, i.e., the mass of $M_\chi=200$~GeV was kept fixed along all calculations. Here we relax this condition and let the dark particle mass be in the range of $50\mbox{ GeV}\leqslant M_\chi\leqslant 500\mbox{ GeV}$. As already pointed out in Sec.~\ref{form}, such interval~\citep{coolingdm} produces spin-independent scattering cross-sections compatible with data provided by some experiments~\citep{panda,lux,darkside}.

\subsection{Mass-radius diagrams and tidal deformability}
\label{resultsmr}

\begin{figure*}
\centering
\includegraphics[scale=0.44]{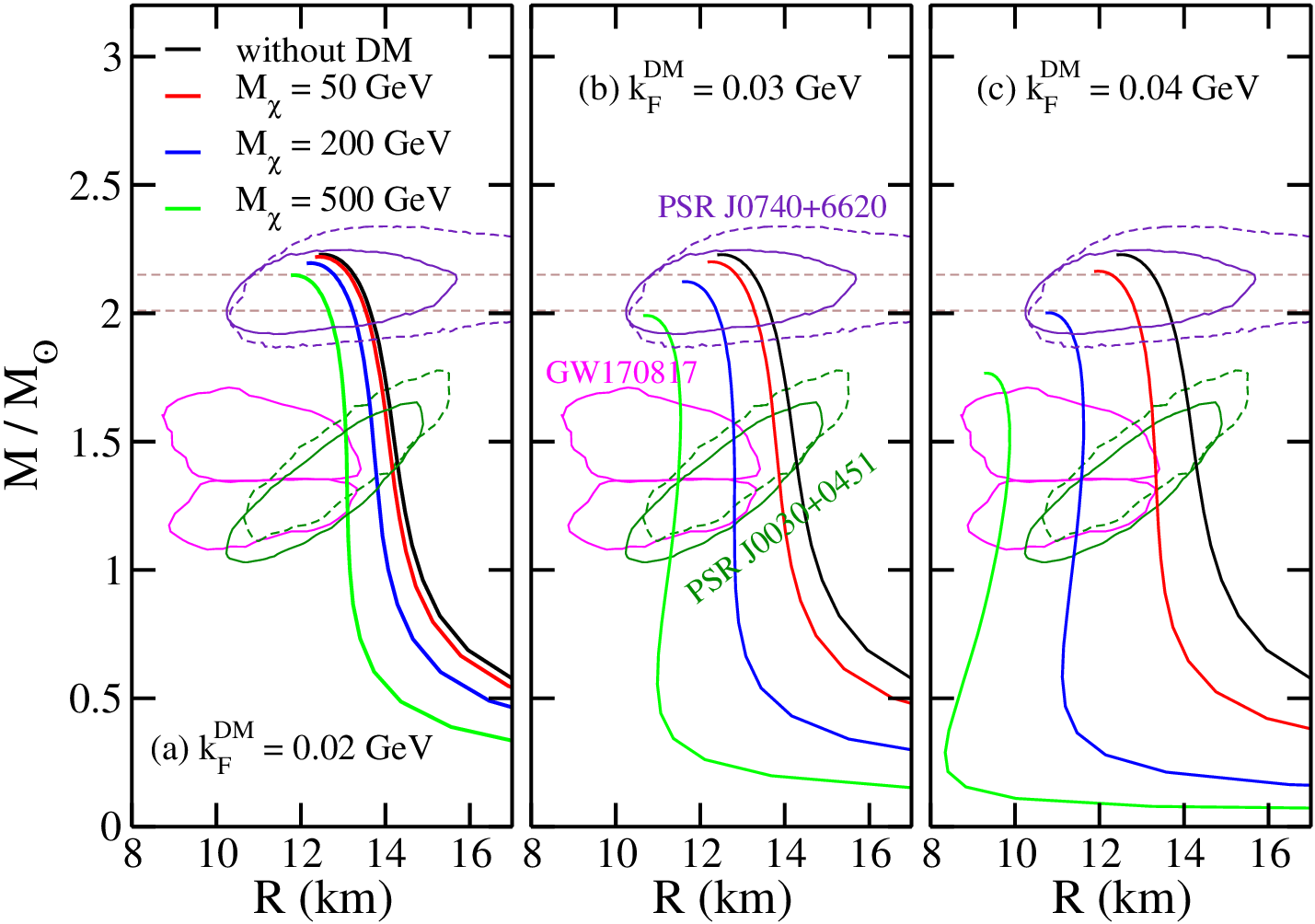}
\caption{Mass-radius diagrams constructed from the \mbox{RMF-SRC-DM} model for different values of $M_\chi$ and $k_F^{\mbox{\tiny DM}}$. Black curves indicate the case in which no DM is included. The contours are related to data from the NICER mission, namely, PSR J0030+0451~\citep{0451a,0451b} 
and PSR J0740+6620~\citep{6620a,6620b}, and the GW170817 event~\citep{ligo17,ligo18}, all of them at $90\%$ credible level. The horizontal lines refer to recent observational constraints on neutron star mass~\citep{fonseca}.} 
\label{mr}
\end{figure*}

In order to construct the mass-radius diagrams for the \mbox{RMF-SRC-DM} model presented in the previous sections, it is needed to solve the Tolman-Oppenheimer-Volkoff (TOV) equations~\citep{tov39,tov39a} given by
\begin{align}
\frac{dp(r)}{dr}&=-\frac{[\epsilon(r) + p(r)][m(r) + 4\pi r^3p(r)]}{r^2[1-2m(r)/r]},
\label{tov1}
\\
\frac{dm(r)}{dr}&=4\pi r^2\epsilon(r).
\label{tov2}
\end{align}
The solution of these equations is constrained to $p(0)=p_c$ (central pressure) and $m(0) = 0$. Moreover, one has $p(R) = 0$ and $m(R)\equiv M$ at the star surface, with $R$ defining the radius of the respective star of mass $M$. The description of the neutron star crust is made by parts, namely, the outer crust, modeled by the Baym-Pethick-Sutherland equation of state~\citep{bps} for $6.3\times10^{-12}\,\mbox{fm}^{-3} \leqslant\rho\leqslant 2.5\times10^{-4}\,\mbox{fm}^{-3}$~\cite{poly2,malik19}, and the inner crust, for which the polytropic expression $p(\epsilon)=A+B\epsilon^{4/3}$~\citep{poly2,poly1,gogny2} is used in the range of $2.5\times10^{-4}\,\mbox{fm}^{-3} \leqslant\rho\leqslant \rho_t$. The transition density, denoted by $\rho_t$, is found from the thermodynamical method~\citep{cc2,kubis04,gonzalez19}. Furthermore, we consider a system composed of protons, neutrons, leptons, and dark matter under charge neutrality and $\beta$-equilibrium. Therefore, the total energy density and pressure used as input for the TOV equations are $\epsilon=\mathcal{E}+\sum_l\epsilon_l$ and $p=P + \sum_lp_l$, with $\mathcal{E}$ and $P$ given in Eqs.~(\ref{eden}) and (\ref{press}), respectively, with $l$ referring to the leptons (muons, with $m_\mu=105.7$~MeV, and massless electrons). {We emphasize that in our approach, the contribution of DM is distributed in the core of the star, since for the outer shell (crust region) we assume the aforementioned construction (piecewise equation of state) in which only ordinary matter is taken into account.}

In Fig.~\ref{mr} we display the mass-radius diagrams generated by the \mbox{RMF-SRC-DM} model for three different values of the dark mass particle, namely, $M_\chi=50$~GeV, $200$~GeV and $500$~GeV, and for its Fermi momentum given by $k_F^{\mbox{\tiny DM}}=0.02$~GeV, $k_F^{\mbox{\tiny DM}}=0.03$~GeV and $k_F^{\mbox{\tiny DM}}=0.04$~GeV. For the sake of comparison, we also provide the curve related to the \mbox{RMF-SRC} model, i.e., without dark matter content in the system. This particular case is constructed by taking $k_F^{\mbox{\tiny DM}}=0$. Rigorously, the term proportional to $h^2$ would still nonvanishing in Eqs.~(\ref{eden}) and~(\ref{press}). However, it was shown in~\cite{dmnosso1} that this particular term is negligible in comparison to the other ones. Therefore, the pure hadronic system is recovered by making $k_F^{\mbox{\tiny DM}}=0$ in the equations of the model. From Fig.~\ref{mr}, it is clear that the inclusion of dark matter content, by taking nonvanishing values of $k_F^{\mbox{\tiny DM}}$, reduces the neutron star mass as previously mentioned. Furthermore, it is also verified that the increase of $M_\chi$ also favors this reduction. This is a feature in agreement with other studies performed in which the dark particle mass is allowed to run, see, for instance, \cite{feeble-vector,feeble,abdul}. Notice that the neutron stars radii are also highly affected by the change of $M_\chi$. In particular, $R$ decreases as $M_\chi$ increases and this variation is larger for higher values of $k_F^{\mbox{\tiny DM}}$. 

We also depict in Fig.~\ref{mr} the limiting contours related to recent observational data obtained (i) from NASA’s Neutron star Interior Composition Explorer (NICER) mission, regarding measurements of pulsars PSR J0030+0451~\citep{0451a,0451b} and PSR J0740+6620~\citep{6620a,6620b}, determined in 2019 and 2021, respectively, (ii) from the LIGO and Virgo Collaboration (LVC) concerning the detection of gravitational waves caused by the coalescence of a binary neutron star system, the so called GW170817 event~\citep{ligo17,ligo18}, and (iii) from a new measurement provided on the maximum neutron star mass~\citep{fonseca} (horizontal dashed lines). As one can observe, it is completely possible to generate \mbox{RMF-SRC-DM} parametrizations, by varying $k_F^{\mbox{\tiny DM}}$ and/or $M_\chi$,  in agreement with all these constraints, i.e, a certain DM content can be used to predict mass-radius diagrams simultaneously consistent with observations.

{For another view of the impact of the neutralino mass on the stellar matter, we show in Fig.~\ref{profile} the profiles related to stars of $1.4M_\odot$ and $1.8M_\odot$.
\begin{figure}
\centering
\includegraphics[scale=0.36]{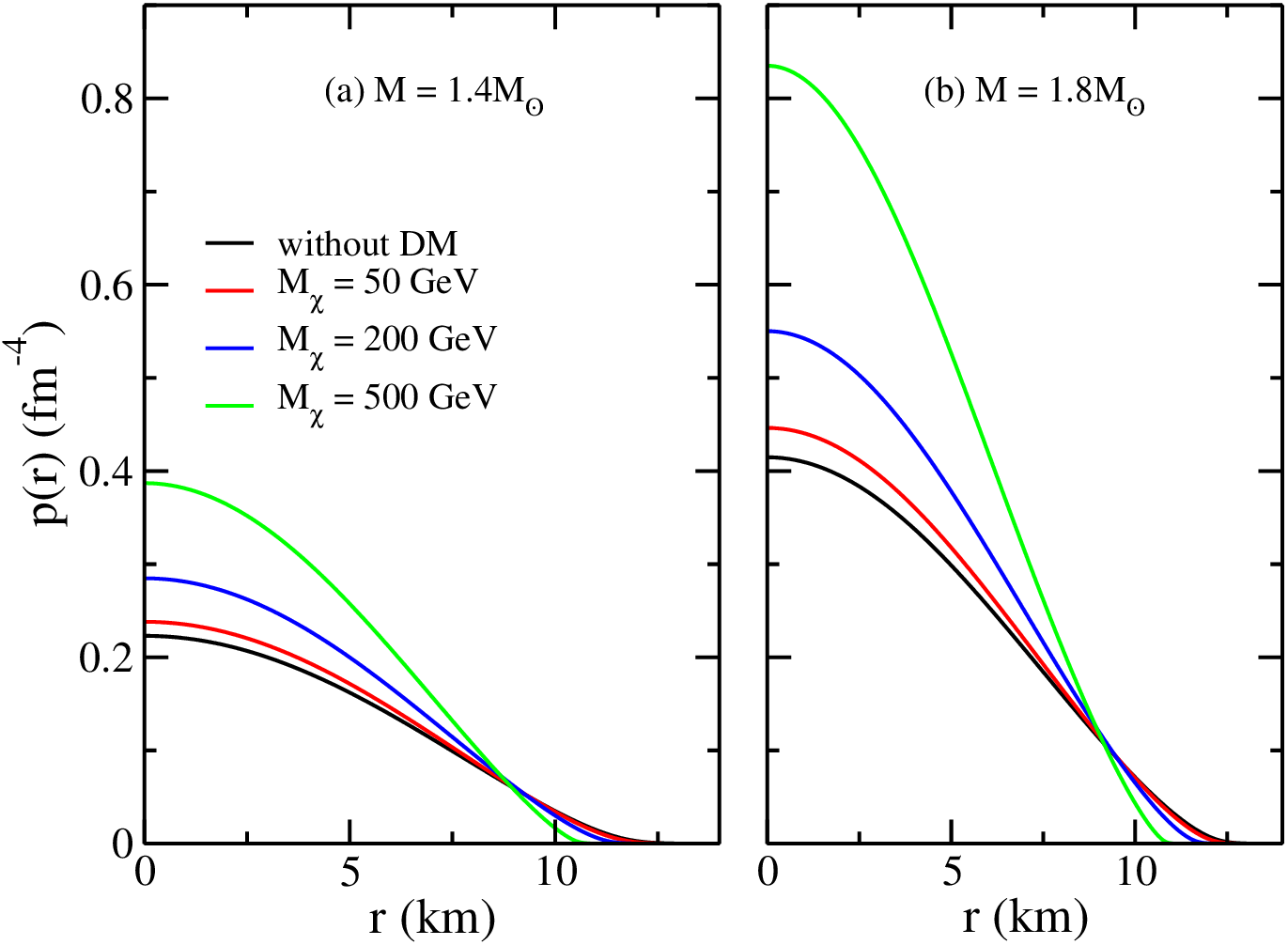}
\caption{{Pressure as a function of the radius for a (a) $1.4M_\odot$, and (b) $1.8M_\odot$ star constructed by taking $k_F^{\mbox{\tiny DM}}=0.03$~GeV and different values of $M_\chi$. Black curves: case in which no DM is included.}}
\label{profile}
\end{figure}
From this figure, it is observed that the central pressure increases for higher values of $M_\chi$, with the variation being more significant for the more massive star. Moreover, notice that the pressure change is more concentrate in the inner region of the star, in agreement with results found in~\cite{coolingdm}. A more detailed investigation on the size of the DM inside the star can be easily performed by adopting the two-fluid approach, in which profiles of each component can be obtained separately and, consequently, precise information regarding DM halos or DM concentrated in the CS core can be extracted. In a recent study performed in~\cite{giangrandi-pos,constanca}, for instance, the authors shown that accumulated DM inside the star favor the compatibility with the GW170817 data.
}

\begin{figure*}
\centering
\includegraphics[scale=0.44]{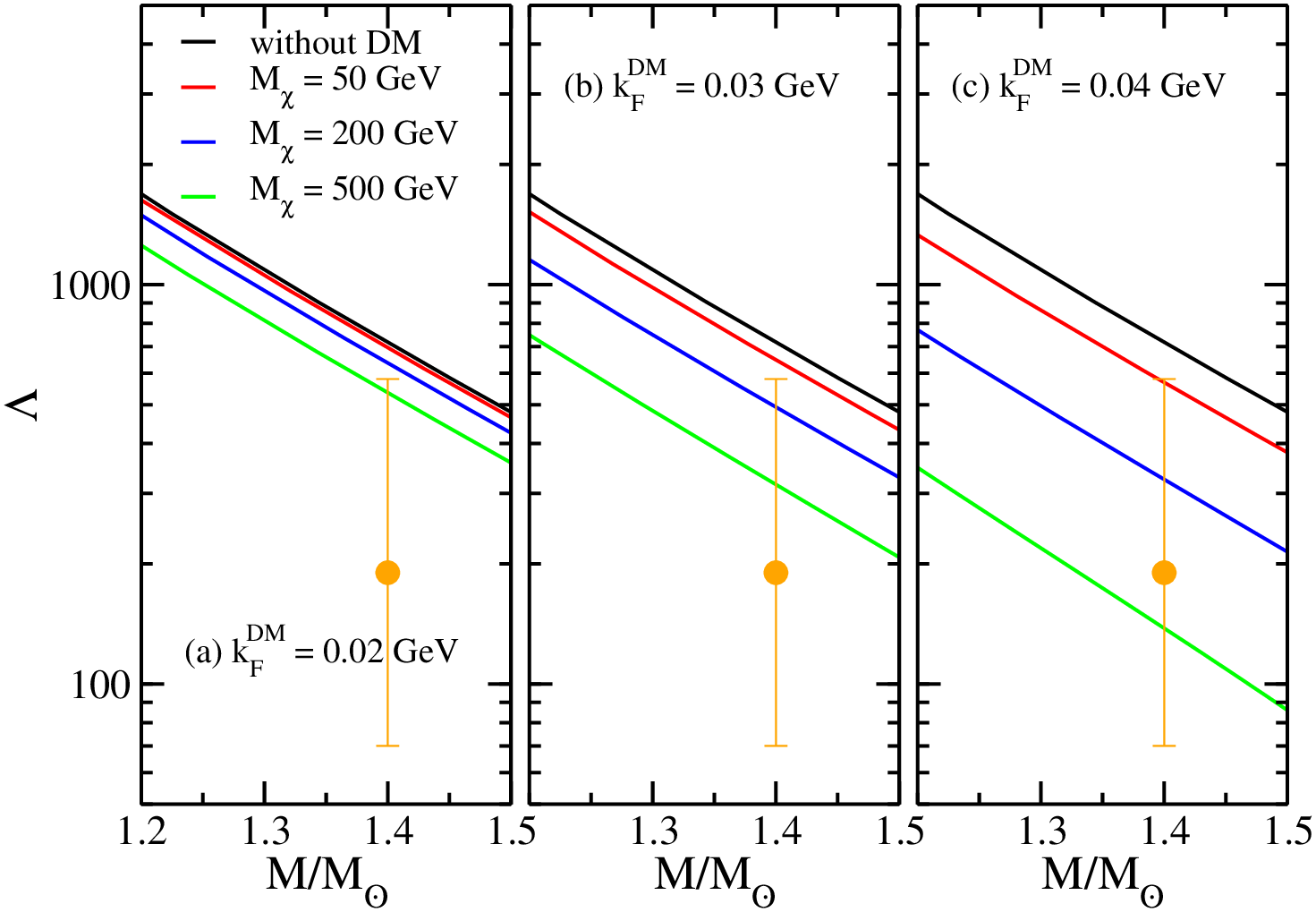}
\caption{$\Lambda$ (log scale) as a function of $M$ (solar mass units) constructed from the \mbox{RMF-SRC-DM} model for different values of $M_\chi$ and $k_F^{\mbox{\tiny DM}}$. Circle with error bars: result of $\Lambda_{1.4}=190^{+390}_{-120}$ determined by LVC~\citep{ligo18}.} 
\label{lambda}
\end{figure*}
\begin{figure*}
\centering
\includegraphics[scale=0.44]{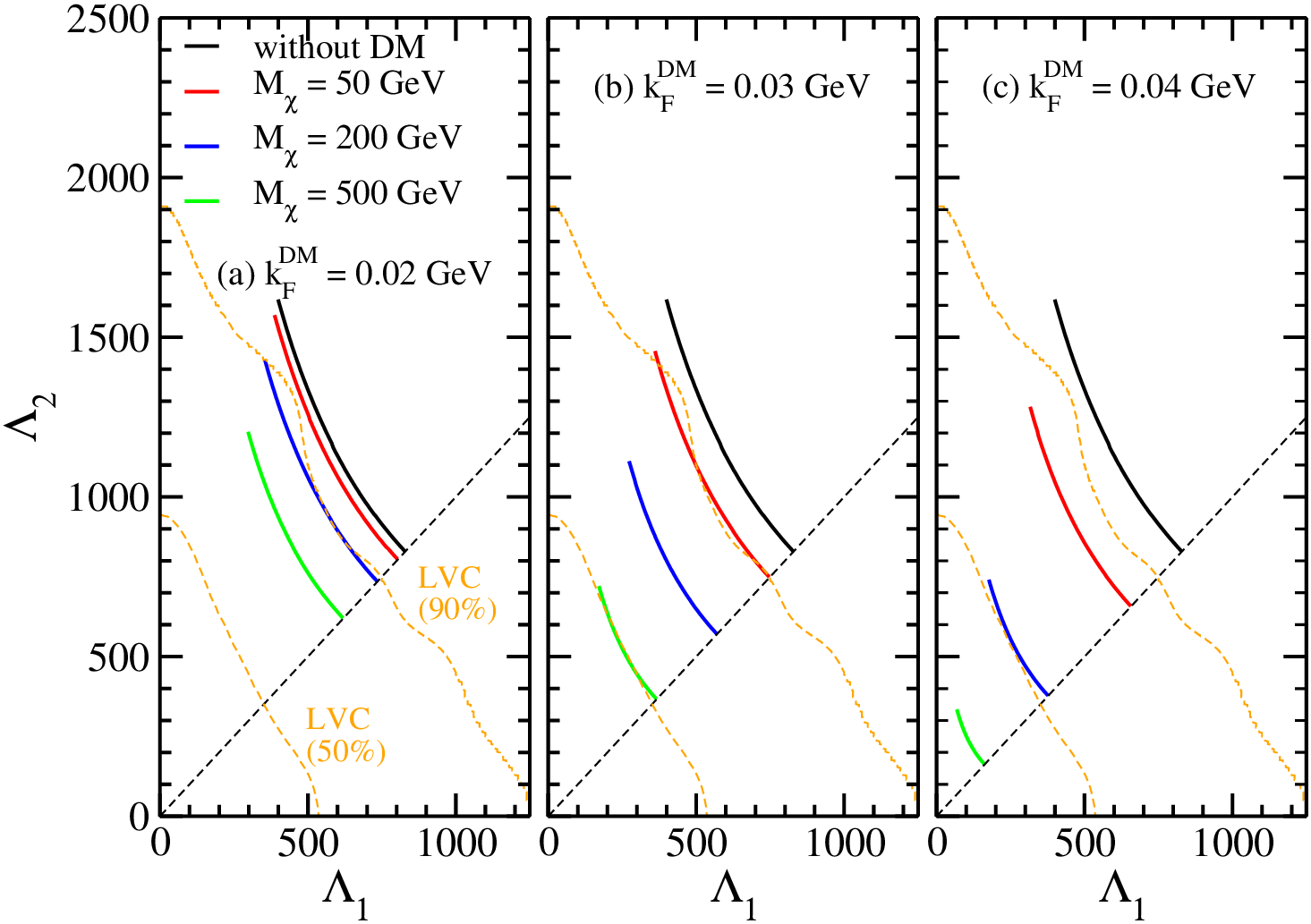}
\caption{Dimensionless tidal deformabilities for the case of high-mass ($\Lambda_1$) and low-mass ($\Lambda_2$) components of the GW170817 event. Results obtained from the \mbox{RMF-SRC-DM} model for different values of $M_\chi$ and $k_F^{\mbox{\tiny DM}}$. Confidence lines, namely, $90\%$ and $50\%$, taken from~\citep{ligo18}.} 
\label{l1l2}
\end{figure*}

In order to obtain the dimensionless tidal deformability, defined as $\Lambda = 
2k_2/(3C^5)$, with $C=M/R$,
\begin{eqnarray}
&k_2& = \frac{8C^5}{5}(1-2C)^2[2+2C(y_R-1)-y_R]\nonumber\\
&\times&\Big\{2C [6-3y_R+3C(5y_R-8)] \nonumber\\
&+& 4C^3[13-11y_R+C(3y_R-2) + 2C^2(1+y_R)]\nonumber\\
&+& 3(1-2C)^2[2-y_R+2C(y_R-1)]{\rm ln}(1-2C)\Big\}^{-1},\qquad
\label{k2}
\end{eqnarray}
and $y_R=y(R)$, we proceed to determine the function $y(r)$ through the solution of the differential equation given by 
\begin{eqnarray}
r\frac{dy}{dr} + y^2 + yF(r) + r^2Q(r) = 0,
\end{eqnarray} 
solved along with the TOV equations. The quantities $F(r)$ and $Q(r)$ read
\begin{eqnarray}
F(r) &=& \frac{1 - 4\pi r^2[\epsilon(r) - p(r)]}{g(r)} , 
\\
Q(r)&=&\frac{4\pi}{g(r)}\left[5\epsilon(r) + 9p(r) + 
\frac{\epsilon(r)+p(r)}{v_s^2(r)}- \frac{6}{4\pi r^2}\right]
\nonumber\\ 
&-& 4\left[ \frac{m(r)+4\pi r^3 p(r)}{r^2g(r)} \right]^2,
\label{qr}
\end{eqnarray}
with $v_s^2(r)=\partial p(r)/\partial\varepsilon(r)$ (squared sound velocity)~\citep{tanj10,postnikov10,hind08,damour,tayl09}. 

In Fig.~\ref{lambda} we show how $\Lambda$ depends on the neutron star mass for different \mbox{RMF-SRC-DM} parametrizations.
From this figure, it is clear that the increase of $M_\chi$ or $k_F^{\mbox{\tiny DM}}$ favors the model to reach the limit of $\Lambda_{1.4}=190^{+390}_{-120}$ provided by LVC~\citep{ligo18}. In particular, the effect of the $M_\chi$ variation is stronger for higher values of the dark particle Fermi momentum, as also verified in the mass-radius diagrams. In Fig.~\ref{l1l2} are displayed the tidal deformabilities $\Lambda_1$ and $\Lambda_2$ of the binary neutron star system (masses $M_1$ and $M_2$) associated to the GW170817 event, with $1.37\leqslant M_1/M_\odot \leqslant 1.60$~\cite{ligo17}.
The mass $M_2$ of the companion star is determined through the relationship between $M_1$, $M_2$ and $\mathcal{M}$ (chirp mass) given by~\citep{ligo17}
\begin{eqnarray}
\mathcal{M} = (M_1M_2)^{3/5}/(M_1+M_2)^{1/5}=1.188M_\odot,
\end{eqnarray}
resulting in $1.17 \leqslant M_2/M_\odot \leqslant 1.36$~\citep{ligo17,ligo18}. The upper and lower orange lines correspond to the 90\% and 50\% confidence limits also related to the GW170817 event and provided by LVC~\citep{ligo18}. From the figure, it is verified that $M_\chi$ and $k_F^{\mbox{\tiny DM}}$ induce the curves to reach the regions defined by the observational constraint given by the LVC. Notice that the ``move'' of the curves in direction to the $50\%$ region is more pronounced for the case in which $M_\chi$ increases for higher values of $k_F^{\mbox{\tiny DM}}$, exactly as in the case of the observation data for $\Lambda_{1.4}$ in Fig.~\ref{lambda}.

\subsection{Crustal properties}
\label{resultscrust}

In this section, we show results provided by the \mbox{RMF-SRC-DM} model concerning the properties related to the neutron star crust~\citep{haensel-review,haenselbook,ravenhall-review,harishcrust,margueroncrust,limcrust}. The outer layer of this astrophysical compact object, with subsaturation densities, contains only a small fraction of the entire system, typically a few percent of the {CS} mass. However, the crust physics is very important for the description of the {CS} phenomenology, namely, cooling, r-processes, oscillations in gamma-ray repeaters, and thermal relaxation in x-ray transients, for instance. In particular, this specific region of a {CS} needs to be modeled in order to correctly compute the coexistence of an electron gas, a lattice of neutron-rich nuclei, and nonspherical shape structures, such as rods, slabs, waffles, and parking garages~\citep{schneider18,newton09,schuetrumpf19,debora08}, appearing in the so called pasta phase~\citep{ravenhall}, namely, a frustrated system in which an intense competition between electromagnetic and strong interactions takes place. For this purpose, some treatments are used to describe the {CS} crust, such as the piecewise approach used to generate the results presented in the last section, in which the crust is divided into outer and inner parts. Another formulation takes into account an unified equation of state used as input to the TOV equations. In this case, the same hadronic parametrization is applied for the description of all {CS} regions including the crust (outer + inner parts). However, this treatment is more difficult to be implemented. Its construction for different kinds of hadronic models is a hard task, see for instance~\cite{gognycrust} for a study in this direction performed from nonrelativistic Gogny interactions. 

\begin{figure*}
\centering
\includegraphics[scale=0.44]{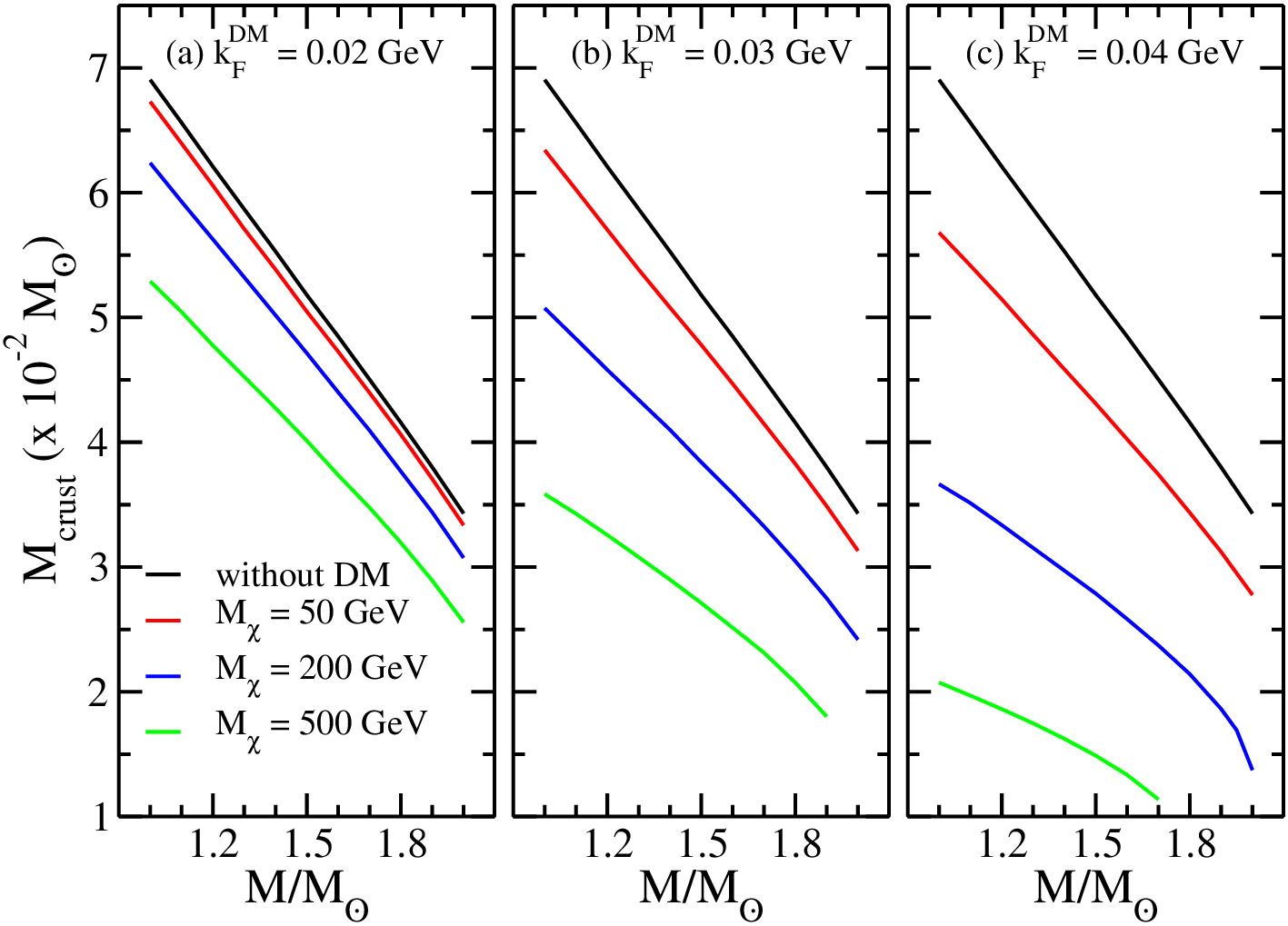}
\caption{Neutron star crustal mass against the total mass $M$ (solar mass units), constructed from the \mbox{RMF-SRC-DM} model for different values of $M_\chi$ and $k_F^{\mbox{\tiny DM}}$. Black curves indicate the case in which no DM is included.} 
\label{mcrust}
\end{figure*}

Due to the crust structure, namely, a very thin system at low density regime, it is often in the literature the use of suitable approximations capable of providing analytical expressions for some quantities related to this part of the {CS}~\citep{link99,lattimer-review1,lattimer01,lorenz93}. In particular, the studies performed in~\cite{piekarewicz10,piekarewicz2014} by using $M(r) \approx M$, $p(r)\ll \epsilon(r)$, and $4\pi r^3p(r)\ll M$, assumed valid in the crust region, gave rise to
\begin{eqnarray}
M_{\mbox{\tiny crust}}(M) \approx 8\pi R_t^3 p_t\left(\frac{R_t}{2M} - 1\right)
\left[1 + \frac{32p_t}{5\epsilon_t}\left(\frac{R_t}{2M} - \frac{3}{4}\right)\right]
\label{eqmcrust}
\end{eqnarray}
for the mass of the {CS} crust, where $R_t$, $p_t$, and $\epsilon_t$ are, respectively, radius, pressure, and energy density at the crust-core interface, found in this work as described in the last section, i.e., from the thermodynamical method. This expression is used to generate Fig.~\ref{mcrust}. 
Notice that the inclusion of DM content clearly induces a decrease in $M_{\mbox{\tiny crust}}$ in both situations: increase of both, $k_F^{\mbox{\tiny DM}}$ and $M_\chi$. Once again, more pronounced changes are verified by varying $M_\chi$ at higher values of~$k_F^{\mbox{\tiny DM}}$.

\begin{figure*}
\centering
\includegraphics[scale=0.44]{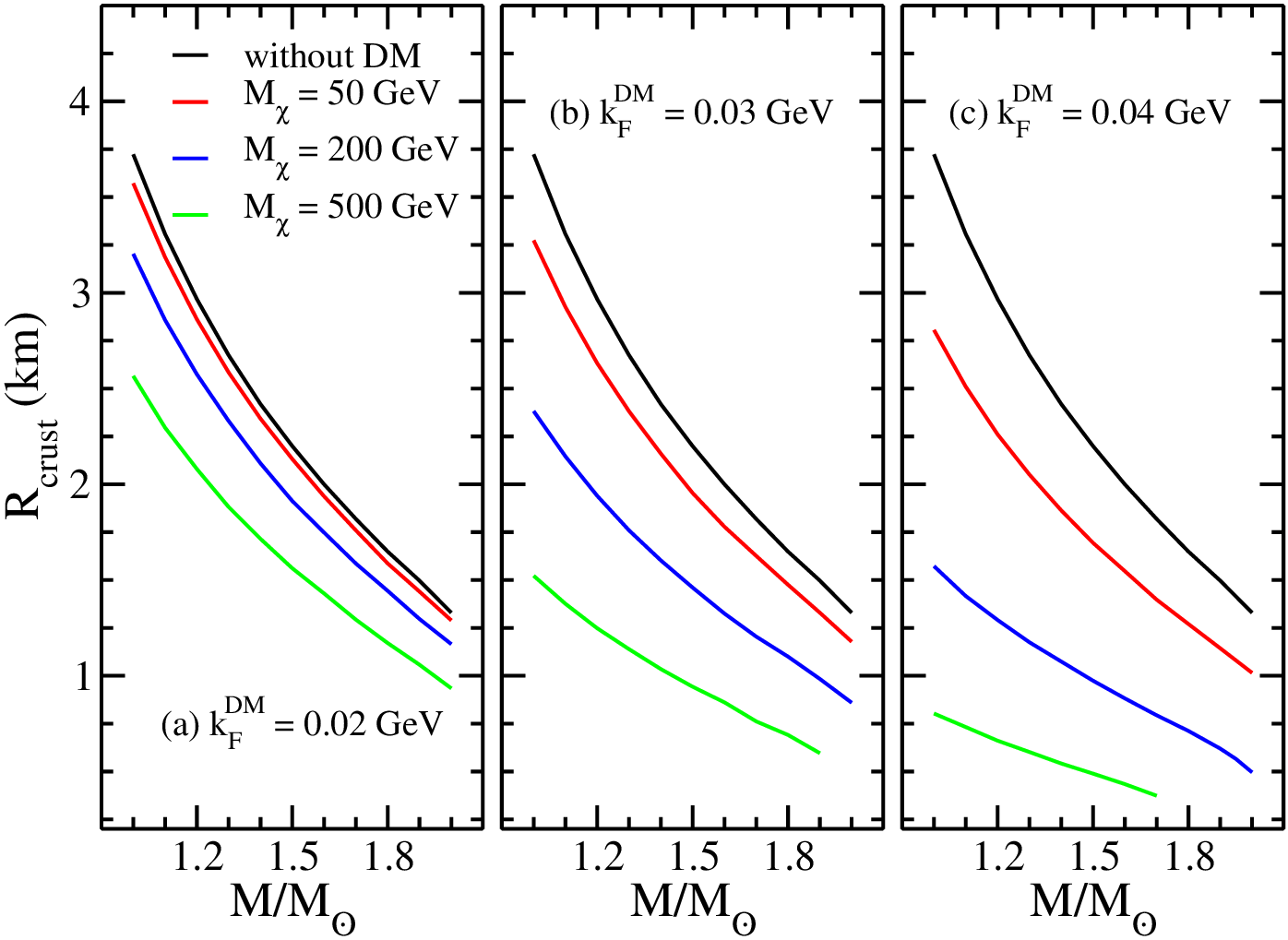}
\caption{Neutron star crustal thickness against the total mass $M$ (solar mass units), constructed from the \mbox{RMF-SRC-DM} model for different values of $M_\chi$ and $k_F^{\mbox{\tiny DM}}$. Black curves indicate the case in which no DM is included.} 
\label{rcrust}
\end{figure*}

Fig.~\ref{rcrust} depicts how the crustal thickness changes as a function of~$M$ for different \mbox{RMF-SRC-DM} parametrizations. 
In order to determine this quantity, we proceed by taking $R_{\mbox{\tiny crust}}=R - R_t$. In this calculation, and also in Eq.~\ref{eqmcrust}, we obtain $R$ by solving the TOV equations as described in the last section and calculate $R_t$ by associating it to the transition point. Notice that the global behavior verified in the previous results is repeated for $R_{\mbox{\tiny crust}}$. The features displayed in Figs.~\ref{mcrust} and~\ref{rcrust} indicate that, for a fixed value of $M$, the increase of dark matter reduces the crustal region since both, mass and thickness, decrease as $k_F^{\mbox{\tiny DM}}$ or $M_\chi$ increases. Actually, such a reduction also happens for the entire star as one can observe in Fig.~\ref{mr}. For instance, a {CS} with $M=1.5M_\odot$ and $k_F^{\mbox{\tiny DM}}=0.04$~GeV (Fig.~\ref{mr}{\color{blue}c}) undergoes a reduction in its radius from $13.3$~km to $9.9$~km as the dark particle mass increases from $50$~GeV to $500$~GeV. Concerning the crustal region, the same variation in $M_\chi$ leads to a change from $1.7$~km to $0.50$~km in the thickness, and the mass decreases from $4.3\times 10^{-2}M_\odot$ to $1.5\times 10^{-2}M_\odot$. 

\begin{figure*}
\centering
\includegraphics[scale=0.44]{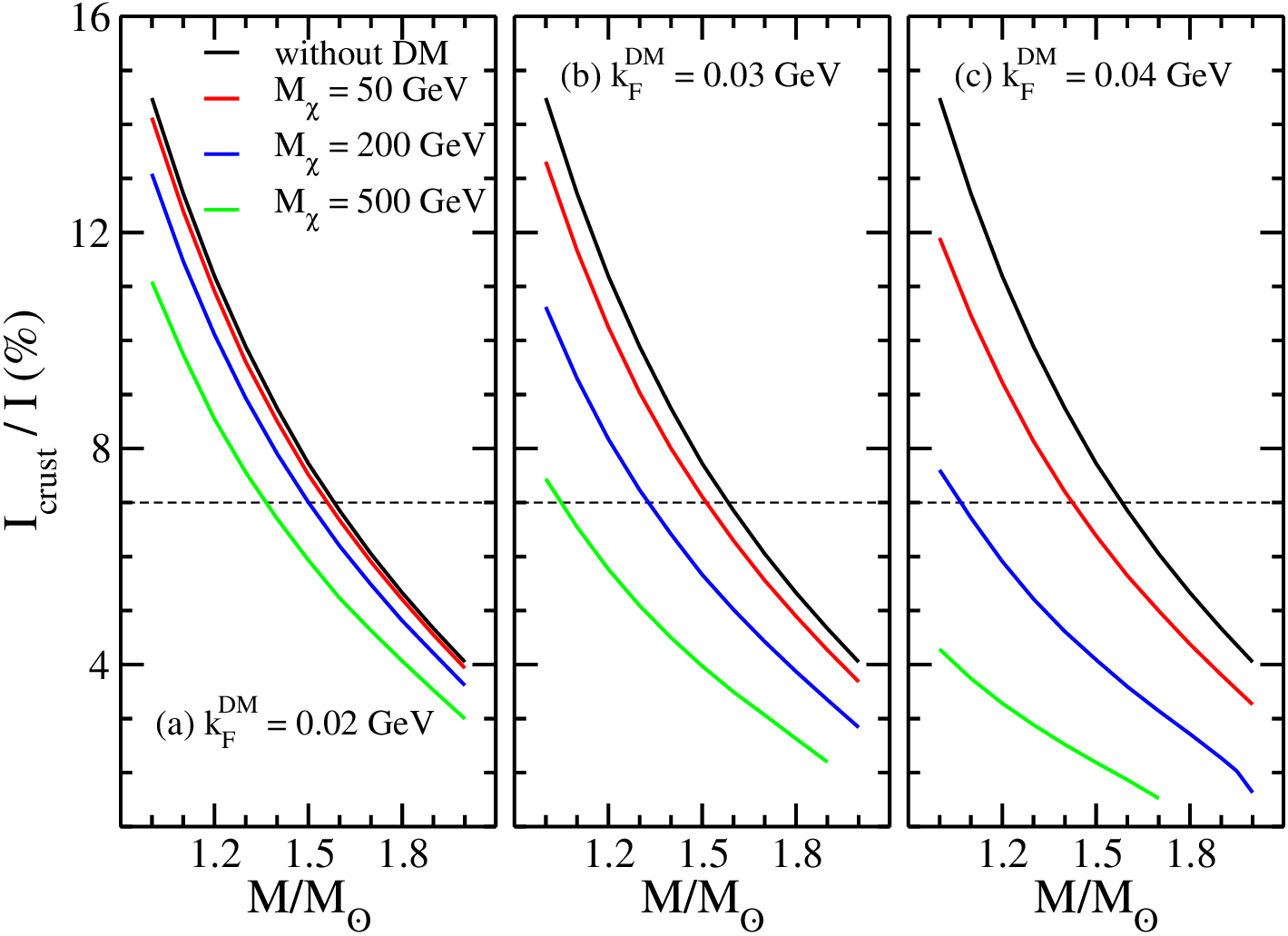}
\caption{Crustal fraction of the moment of inertia as a function of stellar mass for $M$ (solar mass units), constructed from the \mbox{RMF-SRC-DM} model for different values of $M_\chi$ and $k_F^{\mbox{\tiny DM}}$. The horizontal lines indicate the constraint of $7\%$.} 
\label{icrust}
\end{figure*}

Another important phenomenon that can be explained through properties related to the {CS} crust is the suddenly spin-up in the rotational frequency of pulsars~\citep{link99,ho15}, namely, objects that in principle exhibit a very stable rotation, with period from milliseconds to seconds~\citep{hessels06}. Despite present almost atomic clocks precision, the timing behavior of these particular neutron stars can be modified by jumps in their frequencies. These particular interruptions are called glitches~\citep{Lyne-book,Kaspi-2000}, explained from the superfluid vortices exhibited in the neutron star crust~\citep{nature1,nature2}. The difference between the frequencies of the vortices and the entire {CS} can become high enough so the vortices start a move in the outward direction. In order to preserve the conservation of angular momentum, the rotation of the star is increased giving rise to the glitch. It is possible to associate the glitching mechanism to the crustal fraction of the moment of inertia, $I_{\mbox{\tiny crust}}/I$~\citep{Atta-2017,Madhuri-2017,Margaritis}. It was shown in \cite{link99} that glitches can be explained if $I_{\mbox{\tiny crust}}/I > 1.4\%$, namely, value obtained from the analysis of some known glitching pulsars, such as the Vela one. Furthermore, if crustal entrainment effects are also included in the analysis, this constraint changes to~$I_{\mbox{\tiny crust}}/I>7\%$~\citep{PRL109-241103,PRL110-011101}. In order to the capability of the \mbox{RMF-SRC-DM} model to attain this number, we use here the following analytical expression for the crustal moment of inertia,
\begin{align}
I_{\mbox{\tiny crust}}(M) \approx \frac{16\pi R_t^6 p_t}{6M}\left[1 - \frac{0.21}{(R/2M - 1)}\right]
\left[1 + \frac{48p_t}{5\epsilon_t}\left(\frac{R_t}{2M} - 1\right)\right]
\label{eqdeltai}
\end{align}
also extracted from~\cite{piekarewicz10,piekarewicz2014}. The total moment of inertia is obtained from the solution of the Hartle's slow rotation equation, given by~\citep{phil18,hartle,yagi13}, \begin{eqnarray}
0&=&[r-2m(r)]\frac{d^2\omega}{dr^2} - 16\pi r[\epsilon(r)+p(r)]\omega(r)
\nonumber\\
&+& 4\left\{\left(1-\frac{2m(r)}{r}\right) - \pi 
r^2[\epsilon(r)+p(r)]\right\}\frac{d\omega}{dr},
\label{hartle}
\end{eqnarray}
and coupled to the TOV equations. From the function $\omega(r)$, one determines $I=R^3(1-\omega_R)/2$, in which $\omega_R\equiv \omega(R)$. 

In Fig.~\ref{icrust} it is depicted the quantity $I_{\mbox{\tiny crust}}/I$, for the model with dark matter included, alongside the aforementioned observational constraint of $7\%$ for this quantity. 

\begin{figure}
\centering
\includegraphics[scale=0.34]{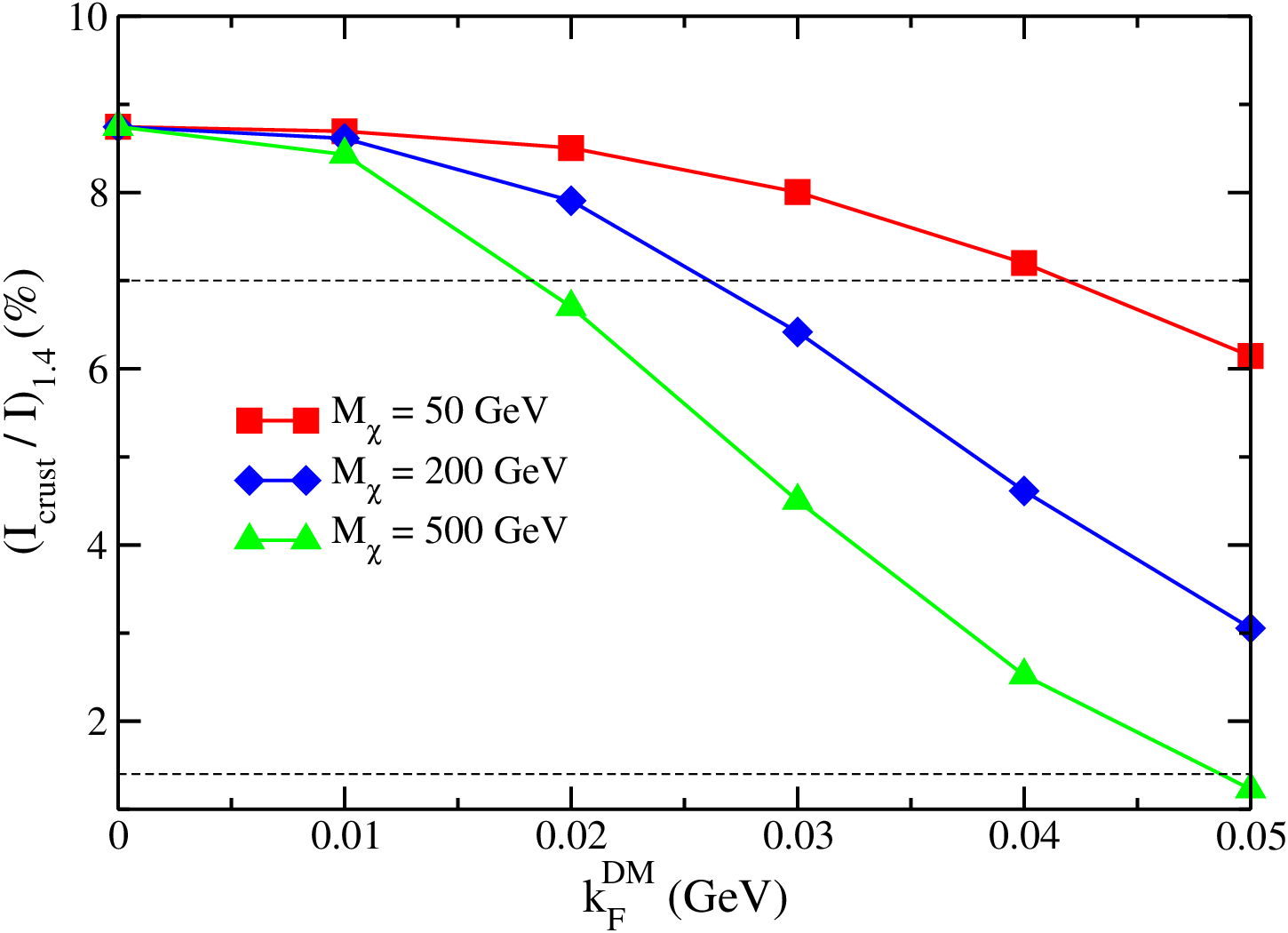}
\caption{$I_{\mbox{\tiny crust}}/I$ for a canonical {CS} of $1.4M_\odot$ as a function of the Fermi dark moment, obtained from the \mbox{RMF-SRC-DM} model for different values of~$M_\chi$. The horizontal lines indicate the constraints of $1.4\%$ and $7\%$.} 
\label{icrust14}
\end{figure}

As one can see, the increase of dark matter content, by increasing the dark Fermi momentum, causes the decrease in the crustal fraction of the moment of inertia for all ranges of {CS} mass investigated. This effect is compatible with the reduction of the crust region previously discussed. The variation in $M_\chi$ also produces the same effect, in this case, more significant for higher values of $k_F^{\mbox{\tiny DM}}$, a feature observed in all results presented before. {It is worth to mention that the effect in $I_{\mbox{\tiny crust}}/I$ observed by the inclusion of DM is due to the variation in both, the total moment of inertia ($I$) and the moment of inertial of the crust ($I_{\mbox{\tiny crust}}$) itself. Both quantities are affected by variations in $k_F^{\mbox{\tiny DM}}$ and $M_\chi$. Furthermore,} it is clear that even DM producing reduction in $I_{\mbox{\tiny crust}}/I$, it is still possible to find parametrizations of the \mbox{RMF-SRC-DM} model in agreement with $I_{\mbox{\tiny crust}}/I>7\%$, i.e., the presence of dark matter can be a possible explanation to the glitch activity in rotation-powered pulsars with exhibiting entrainment effects. In order to verify more clearly this situation, we present in Fig.~\ref{icrust14} the crustal fraction of the moment of inertia for a canonical $1.4M_\odot$ {CS}. The value $M=1.4M_\odot$ was adopted for the Vela pulsar in~\cite{pavlov} to fit its x-ray spectrum. However, it is worth remarking that this is an estimation and not a real measurement of the Vela mass (not precisely known). Other estimations are found in literature, for instance in~\cite{ho15} in which the value $M=(1.51\pm 0.04)M_\odot$ was used. 
From the figure it is verified that all parametrizations are compatible with $I_{\mbox{\tiny crust}}/I > 1.4\%$ regardless the value of $M_\chi$ or $k_F^{\mbox{\tiny DM}}$, i.e., there is always some DM content capable of describing the glitching mechanism with no entrainment effects for a pulsar of $1.4M_\odot$. If the constraint analyzed is the $I_{\mbox{\tiny crust}}/I>7\%$ one, then it is still possible to construct a parametrization with $50\mbox{ GeV}\leqslant M_\chi\leqslant 500\mbox{ GeV}$, however, higher values of $M_\chi$ must be followed by lower values of~$k_F^{\mbox{\tiny DM}}$ in order to satisfy the condition. 

\section{Summary and concluding remarks} 
\label{summary}

In this work, we extended the studies performed in~\cite{dmnosso1,dmnosso2}, in which a nucleonic short-range correlated relativistic hadronic model coupled with dark matter was used to describe neutron star properties. Here we used the same structure of the refereed \mbox{RMF-SRC-DM} model and allowed the variation of the dark particle in the range of $50\mbox{ GeV}\leqslant M_\chi\leqslant 500\mbox{ GeV}$, in agreement with data of spin-independent scattering cross-sections obtained from PandaX-II~\citep{panda}, LUX~\citep{lux}, and DarkSide~\citep{darkside} collaborations. 

We have shown that it is possible to find parametrizations with dark matter content capable of simultaneously attaining observational data, in the mass-radius diagram, regarding NASA’s Neutron star Interior Composition Explorer (NICER) mission, the GW170817 event detected by the LIGO and Virgo collaboration (LVC), and recent data on the neutron star mass provided by~\cite{fonseca}. Our findings also pointed out the feature that the increase of $M_\chi$ favors the model to attain the data of $\Lambda_{1.4}=190^{+390}_{-120}$ determined by LVC~\citep{ligo18}, and the ones in the $\Lambda_1\times\Lambda_2$ region related to the neutron stars of the binary system.  

Furthermore, we also calculated some crustal properties, namely, mass ($M_{\mbox{\tiny crust}}$) , thickness ($R_{\mbox{\tiny crust}}$) and crustal fraction of the moment of inertia ($I_{\mbox{\tiny crust}}/I$). We verified that the increase of $M_\chi$ also produces a reduction of the neutron star crust, i.e., $M_{\mbox{\tiny crust}}$ and $R_{\mbox{\tiny crust}}$ decrease in this situation. Concerning $I_{\mbox{\tiny crust}}/I$, it was shown that this quantity decreases as a function of both, $M_\chi$ and the Fermi momentum of the dark particle. However, we verified that it is possible to find \mbox{RMF-SRC-DM} parametrizations that satisfy the constraints $I_{\mbox{\tiny crust}}/I>1.4\%$ and $I_{\mbox{\tiny crust}}/I>7\%$, both related to the glitching mechanism of rotation-powered pulsars, such as the Vela one. The latter is obtained when entrainment effects are taken into account in the glitch activity. As a direct consequence, we verified that dark matter can also be used to help describe the glitch phenomenon in pulsars.

\section*{ACKNOWLEDGMENTS}
This work is a part of the project INCT-FNA proc. No. 464898/2014-5. It is also supported by Conselho Nacional de Desenvolvimento Cient\'ifico e Tecnol\'ogico (CNPq) under Grants No. 312410/2020-4 (O.L.), and No. 308528/2021-2 (M.D.). We also acknowledge Funda\c{c}\~ao de Amparo \`a Pesquisa do Estado de S\~ao Paulo (FAPESP) under Thematic Project 2017/05660-0 and Grant No. 2020/05238-9.

\section*{DATA AVAILABILITY STATEMENT}
This manuscript has no associated data or the data will not be deposited. All data generated during this study are contained in this published article.

% \newpage
% \bibliographystyle{apssamp}
\bibliographystyle{mnras}
\bibliography{references-revised.bib}

\end{document}